\documentclass[journal]{IEEEtran}
 \pdfoutput=1
\ifCLASSINFOpdf
\usepackage{listing}
  \usepackage[pdftex]{graphicx}
	\usepackage[pdftex]{hyperref}
	\usepackage{booktabs}
	\usepackage{amsmath}
	\usepackage{slashbox}
	\usepackage{textcomp}
	\usepackage{multirow}
	\usepackage{textcomp}
	\usepackage[euler]{textgreek}

\usepackage[hang,font=small]{caption}%
	\usepackage{color,soul}
	
	\definecolor{gray}{rgb}{1,0.6,0.5}
   \usepackage[square, comma, sort&compress,numbers]{natbib}
   \graphicspath{{./figures/}}
\else
\fi
\allowdisplaybreaks

\begin{document}
\title{ViSTRA2: Video Coding using Spatial Resolution and Effective Bit Depth Adaptation}

\author{Fan~Zhang,~\IEEEmembership{Member,~IEEE},
~Mariana~Afonso,~\IEEEmembership{Member,~IEEE},
        and ~David~R.~Bull,~\IEEEmembership{Fellow,~IEEE}
\thanks{Manuscript drafted at September 2019.} 
\thanks{Fan Zhang, Mariana Afonso, and David R. Bull are with the Visual Information Laboratory, University of Bristol, Bristol, UK.}
\thanks{E-mail: fan.zhang@bristol.ac.uk, mariana.afonso@bristol.ac.uk, and dave.bull@bristol.ac.uk}}
\maketitle

\begin{abstract}

We present a new video compression framework (ViSTRA2) which exploits adaptation of  spatial resolution and effective bit depth, down-sampling these parameters at the encoder based on perceptual criteria, and up-sampling at the decoder using a deep convolution neural network. ViSTRA2 has been integrated with the reference software of both the HEVC (HM 16.20) and VVC (VTM 4.01), and evaluated under the Joint Video Exploration Team Common Test Conditions using the Random Access configuration. Our results show consistent and significant compression gains against HM and VVC based on Bj{\o}negaard Delta measurements, with average BD-rate savings of 12.6\% (PSNR) and 19.5\% (VMAF) over HM and 5.5\% (PSNR) and 8.6\% (VMAF) over VTM.

\end{abstract}


\IEEEpeerreviewmaketitle

\section{Introduction}
\label{sec:intro}

The development of video technology for film, television, terrestrial and satellite transmission, surveillance, and especially Internet video, demands much higher bandwidth than ever before to support new services and ever increasing numbers of users consuming ever more content. Although communication techniques have improved  significantly recently due to the advances in network and physical layers, the bitrate available to an individual user at the application layer is still limited, due to  the increasing amounts of video data (with higher quality and more immersive formats) consumed everyday. Video compression offers the solution to this problem, but equally presents major challenges in how we achieved yet higher coding gains.  

Since the first video coding standard, H.120 \cite{r:h120}, was introduced in 1984, standardisation of video formats and coding algorithms have played an important role in the application and success of video technology \cite{b:Bull}. In the last three decades, video compression standards have been improved significantly, with each new standard providing approximately twice the compression efficiency compared to its predecessor. The latest effort, Versatile Video Coding (VVC) \cite{s:VVC1}, initiated in 2018, is targeting 30-50\% overall bitrate savings over the current standard, High Efficiency Video Coding (HEVC), through integrating more sophisticated tools.

Inspired by recent advances in artificial intelligence, machine learning-based methods have seen increasing utility in video compression algorithms both for end to end compression and to enhance conventional coding tools \cite{j:Liu3,j:Ma1,j:ZhangYun}. Although some of these works, particularly those using Convolutional Neural Networks (CNNs), have reported evident coding gains, very few of them have been adopted by standards due to their high complexity and unconventional architectures needed for the CNN models \cite{s:JVET-Jnotes}. 

Alongside conventional coding tools, spatial and temporal resolution adaptation have also been employed in video compression to improve coding efficiency. This type of approach has previously been adopted for relatively low bitrate scenarios \cite{c:Nguyen,j:Georgis}, as the quality of up-sampled video content can be inconsistent at higher  bitrates  due to the blurring or aliasing artefacts introduced by resolution adaptation. It is also noted that, in these approaches, only spatial and temporal (frame rate) resolution adaptations have been exploited for video compression and there is little reported work on bit depth adaptation.  

Based on our previous work on spatial resolution (SR) and effective bit depth (EBD)  \cite{j:Zhang9,c:Zhang19,c:Zhang23,c:Zhang29,c:Zhang26,p:Zhang} adaptations, and perceptual quality assessment \cite{c:Zhang20}, a new video compression framework, ViSTRA2, is proposed in this paper. It dynamically down-samples the spatial resolution and bit depth of the input video at the encoder, and reconstructs its original resolution and bit depth during decoding. To improve the final reconstruction quality, a deep CNN is employed for both SR and EBD up-sampling,  trained on a large, multi-resolution video database for HEVC and VVC compression at various quantisation levels. The proposed approach has been integrated with the reference test models of HEVC (HM 16.20) and VVC (VTM 4.01), and evaluated on the Joint Video Exploration Team (JVET) Common Test Conditions (CTC). The results exhibit significant coding gains over both  HEVC (HM) and VVC (VTM), with average BD-rate gains (based on PSNR) of 12.6\% and 5.5\% respectively.

The main contributions of this paper are summarised as follows:

\begin{enumerate}
	\item Integration of both spatial resolution (SR) and effective bit depth (EBD) adaptation into one coding framework.
	\item The employment of a single deep CNN for both SR and EBD up-sampling (using different model parameters for various scenarios);
	\item Robust spatial resolution adaptation decisions based on a bespoke SR-dependent quality metric.
	\item Demonstration that using deep networks in an end to end system offers flexibility in the distribution of complexity across that system.
\end{enumerate}

Comparing to this paper, our previous contributions \cite{c:Zhang19,c:Zhang23,c:Zhang29,c:Zhang26} focus solely on the adaptation of spatial resolution or effective bit depth. In addition, the employed CNN architecture in this paper has been enhanced compared to that  in \cite{j:Zhang9,c:Zhang19,c:Zhang23}. Furthermore, an extended training database has been employed here to achieve improved reconstruction results. Comprehensive evaluation results are presented for ViSTRA2, against both HEVC and VVC using various quality assessment methods. An early version of ViSTRA2 was contributed by the University of Bristol (JVET-J0031) to the JVET ``Call for Proposals'' \cite{s:Zhang1} for Versatile Video Coding (VVC).

The remainder of this paper is organised as follows. Section \ref{sec:background} reviews the state-of-the-art of video coding algorithms, specifically covering the approaches based on resolution adaptation and deep learning. In Section \ref{sec:background}, the proposed coding framework is described in detail alongside its important components. The performance of the presented work is fully evaluated in Section \ref{sec:results}, while Section \ref{sec:conclusion} provides the conclusions and outlines future work.

\section{Background}
\label{sec:background}

This section is divided into three subsections. The first overviews the development of video compression standards and the recent advances on royalty-free coding formats. Research work on machine learning based video coding is then briefly reviewed, followed by a summary of compression algorithms employing resolution adaptation.

\subsection{Video Coding Standards: A Brief Overview}

 Since the early 1980s, video compression has been subject to global standardisation. Each standard defines the bitstream format and the decoder structure, alongside the associated test model (reference encoder), which generates standard-compliant streams and provides benchmark coding performance.  

One of the most successful examples, H.264/AVC (Advanced Video Coding) \cite{r:h264} was jointly delivered in 2004 by ITU-T (VCEG) and ISO/IEC (MPEG). It remains the most widely adopted coding standard for Internet streaming, HDTV and Blu-ray players, although its successor H.265/HEVC \cite{r:HEVC},  developed in 2013, offers nearly double the coding efficiency. More recently, aiming to achieve even higher compression performance (30\%-50\%) than HEVC, a new coding standard, Versatile Video Coding (VVC) \cite{s:VVC1} is under development, offering better support for immersive video including high dynamic range and 360\textdegree.   

In order to provide open source and royalty-free solutions for media delivery, the Alliance for Open Media (AOM) \cite{w:AOM} was formed in 2015 by a consortium of companies.  Its first generation codec, AV1 (AOMedia Video 1) \cite{w:AV1}, was released in 2018, and is considered one of the primary competitors of MPEG standards. Other notable coding standards also include Essential Video Coding/MPEG-5 \cite{r:EVC}, Microsoft WMV (Windows Media Video) \cite{w:WMV9}, BBC (British Broadcasting Corporation) Dirac \cite{w:Dirac}, and AVS standards \cite{w:AVS}.  
    
\subsection{Machine Learning based Video Compression}

With recent advances in computational equipment and graphics processing (GPU) devices, deep neural networks, especially convolutional neural networks (CNNs), now offer tractable solutions to many image processing problems. They are being increasingly applied in image and video compression to enhance various coding tools including intra prediction \cite{c:Laude,j:Yeh1,j:Li4,c:Pfaff}, motion estimation \cite{j:Liu2,c:ZhangHan,j:Zhao1,j:Yan}, transforms \cite{c:Liu1}, quantisation \cite{c:Alam}, entropy coding \cite{c:Song,c:Puri} and loop filtering \cite{c:Kuanar,c:Park,j:Yang4,c:Yang}. Detailed reviews on machine learning based compression can be found in \cite{j:Liu3,j:Ma1,j:ZhangYun}. Among the responses to the Call for Proposals for VVC \cite{s:beyondHEVC}, there were five proposals containing coding tools based on neural networks, but few of these have been adopted by VVC \cite{s:VVC1}, though its development is still ongoing. This is mainly due to their relatively high computational complexity, especially when CNN-based calculation has to be executed at the decoder.

\subsection{Resolution Adaptation for Compression}

In video compression, resolution adaptation based approaches are utilised to trade off the relationship between quantisation and (spatial and temporal) resolutions. These methods encode a lower resolution version of a video, and reconstruct its original resolution during decoding. This process can be applied at different coding stages -- for each macroblock or Coding Tree Unit (CTU) \cite{b:Uslubas,c:Nguyen,j:Li}, frame \cite{j:Shen,c:Zhang19}, group of pictures \cite{j:Wang4,j:Zhang9}, or sequence \cite{j:Georgis,j:Dong2}. The reconstruction quality is highly content dependent and relies on the up-sampling methods employed. This is why this type of approach was mainly applicable in low bitrate cases \cite{j:Shen,j:Georgis,c:Nguyen} or for intra coding \cite{c:Zhang19}, when simple interpolation filters were employed. Inspired by the recent development of (especially CNN-based) super-resolution algorithms \cite{c:Ledig,j:Dong,c:Kim,c:Lim}, it is now possible to extend these approaches to higher bitrate ranges with improved reconstruction quality. 

Due to the content-dependent nature of these adaptation approaches, it is also important to characterise the relationship between spatial resolution, quantisation and visual quality. For this purpose, subjective video databases \cite{c:Zhang21,j:Bampis}, were developed to investigate perceptual quality across a range of spatial resolutions, quantisation levels and up-sampling methods. Based on these databases, objective quality metrics have been designed for generic \cite{j:Bampis,w:VMAF} or bespoke \cite{c:Zhang20} applications, which can facilitate quantisation resolution optimisation during resolution adaptation inside the coding loop.  

\section{Proposed Algorithm}
\label{sec:algorithm}

\begin{figure*}[htbp]
\centerline{\includegraphics[width=1.02\linewidth]{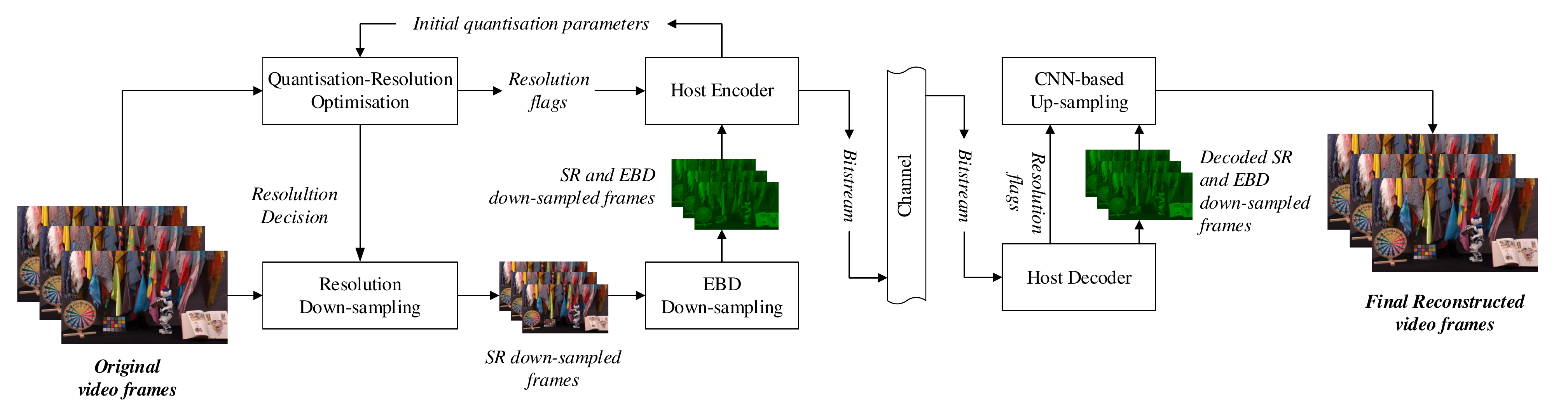}}
\caption{Diagram of the ViSTRA2 encoder. }
\label{fig:workflow}
\end{figure*}

The proposed ViSTRA2 coding framework is illustrated in Fig. \ref{fig:workflow}. At the encoder, the spatial resolution (SR) is firstly determined by a Quantisation-Resolution Optimisation (QRO) module according to the input content and the initial base quantisation parameter ($\mathrm{QP}_\mathrm{base}$) values configured in the host encoder. The original video frames are then spatially down-sampled (by 2 in this case\footnote{Greater improvement may be achieved by applying multiple re-sampling ratios as in \cite{c:Zhang23}. This will however requires more sophisticated CNN training process and accurate QRO prediction. In this paper, as effective bit depth adaptation has also been employed in the coding workflow, a single spatial resolution re-sampling ratio is used for algorithm simplification.}) if SR adaptation is enabled, by applying an anti-aliasing modified separable Lanczos3 filter \cite{b:Turkowski} with a kernel width of 12. In order to signal the SR adaptation decision to the decoder, a flag bit is inserted into the bitsteam as side information.

Secondly, the effective bit depth (EBD) (for both luma and chroma channels) is down-sampled by 1 bit prior to encoding through bitwise right shift. Here EBD is defined as the actual bit depth used to represent the video content, which is different from the coding bit depth (CBD) defined as the pixel bit depth, e.g. \textit{InternalBitDepth} in HEVC HM and VVC VTM reference encoders. Throughout the coding process, EBD is lower than or equal to CBD in the proposed coding workflow, while CBD remains constant. It is noted that there is no optimisation module for EBD adaptation here. This is because, through observing the coding results on the training data and the preliminary results generated in \cite{c:Zhang26}, in most cases, CNN-based EBD adaptation can provide improved (or equivalent) compression performance. Therefore, EBD adaptation is always enabled in the coding framework of ViSTRA2, and does not require any side information indicating the decoder the EBD changes.

During encoding,  to obtain similar bitrate ranges and facilitate comparison with anchor codecs (with the same $\mathrm{QP}_\mathrm{base}$ values), a fixed quantisation parameter (QP) offset is applied on $\mathrm{QP}_\mathrm{base}$. This offset equals -6 when only EBD adaptation is enabled, and becomes -12 if both SR and EBD adaptations are applied. These two values were empirically obtained through the observation on the coding statistics of training sequences \cite{c:Zhang19,c:Zhang26}. 

At the decoder, based on the value of the flag bit, decoded SR and EBD down-sampled video frames are up-sampled to their original SR and EBD using a deep CNN. If SR adaptation is enabled, the spatial resolution of the decoded video frames is firstly up-sampled by 2 using a nearest neighbour filter before CNN reconstruction\footnote{It is noted that, in our previous work \cite{j:Zhang9,c:Zhang23,s:Zhang1}, pre-CNN up-sampling was conducted using a Lanczos3 filter. However we have found that by using a nearest neighhour filter here the overall reconstruction performance can be slightly improved.\label{fn:nn}}. The network architecture and training process alongside the detail description of the QRO module are presented below. 

\subsection{The Employed CNN Architecture}

\begin{figure*}[htbp]
\centerline{\includegraphics[width=1\linewidth]{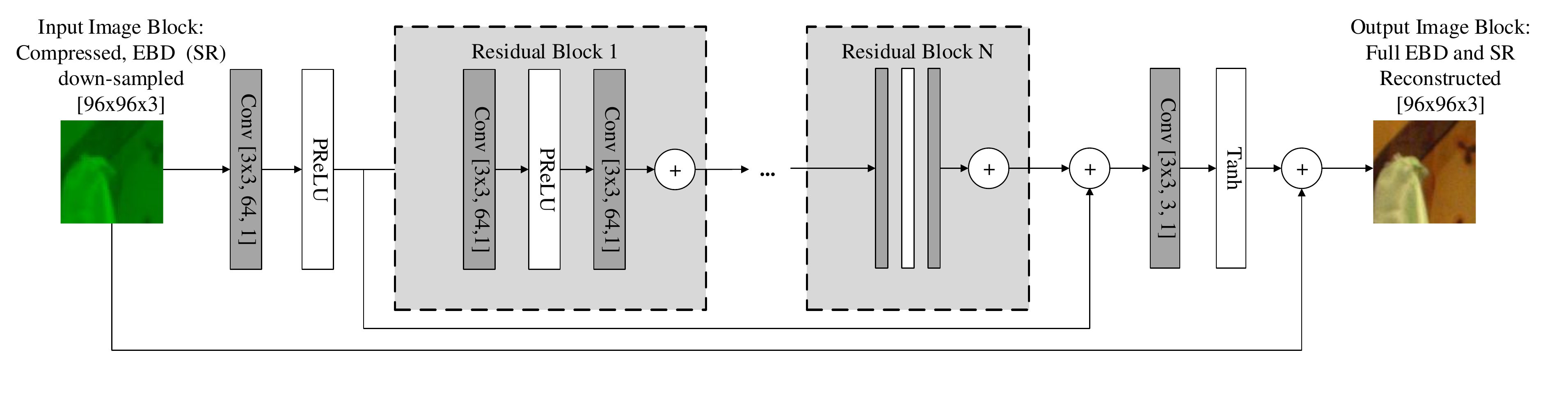}}
\caption{CNN architecture employed for EBD (and SR) up-sampling. }
\label{fig:network}
\end{figure*}

The architecture of the convolutional neural network employed to reconstruct full spatial resolution and effective bit depth is shown in Fig. \ref{fig:network}. The input of the CNN is a 96$\times$96 compressed RGB colour image block with reduced EBD (and re-sampled SR if SR adaptation is enabled), while the output is expected to be the corresponding original image block with the same size. This architecture is modified based on the generator (SRResNet) of SRGAN \cite{c:Ledig}, which was developed for uncompressed image super-resolution. It starts from an initial convolutional layer with a Parametric Rectified Linear Unit (PReLU) as the activation function, and ends with another single convolutional layer with a Tanh activation. Before the final output, a skip connection is employed between the input of the CNN and the output the last convolutional layer. Between these two convolutional layers, there are $N$ identical residual blocks, one of each consisting of two convolutional layers and one PReLU in between. In each residual block, a skip connection is applied between the input of first convolutional layer and the output of the second. Another skip connection is also employed between the output of the first convolutional layer and the output of all residual blocks. In all convolutional layers, the stride value is 1 and the kernel size equals 3$\times$3. The number of feature maps is 64 for most convolutional layers except the last one (3 feature maps there).

Comparing to the original architecture of SRResNet \cite{c:Ledig}, two modifications were applied:

\begin{itemize}
	\item Batch normalisation (BN) layers have not been used here, as they were found to decrease image feature variability and influence overall performance \cite{c:Lim}.
	\item The loss function employed for training the network is $\ell$1 loss rather than $\ell$2 in \cite{c:Ledig}. This is based on recent work on CNN super-resolution \cite{c:Johnson}, which claims improved reconstruction quality achieved from this change. 
\end{itemize}
 
\subsection{Network Training and Evaluation}
\label{sec:training}

The CNN employed for EBD and SR-EBD (if SR adaptation is enabled) up-sampling was trained using 432 video clips at various spatial resolutions (108 source sequences $\times$ 4 spatial resolutions), including 3840$\times$2160, 1920$\times$1080, 960$\times$540, 480$\times$270, each of them having 64 frames with a CBD (coding bit depth) of 10 bits. All clips were collected from publicly available databases, including BVI-HFR \cite{j:Zhang8}, BVI-Texture \cite{c:Zhang10}, Netflix Chimera \cite{w:NetflixChimera} and Harmonic 4K \cite{w:Harmonic}. Their EBD (by 1 bit) and SR-EBD down-sampled (SR adaptation ratio of 2) versions were compressed using both HEVC HM 16.20 and VVC VTM 4.01 encoders for five different initial base QP values 22, 27, 32, 37 and 42 (QP offsets of -6 or -12 was applied during encoding for EBD or SR-EBD version). The same coding configuration was used as the Random Access (RA) mode (Main10 profile) in the Joint Video Exploration Team (JVET) Common Test Conditions (CTC), with a fixed intra period of 64. 

This results in 20 groups of reconstructed videos (2 codecs $\times$ 2 adaptation versions $\times$ 5 QP groups), each group containing 432 reconstructed sequences for a specific codec (HM or VTM), adaptation version (EBD or SR-EBD) and QP group (22-42). For each SR-EBD down-sampled and compressed video, its frames were spatially up-sampled to their original spatial resolution using a nearest neighbour filter (see footnote \ref{fn:nn}). The video frames of all reconstructed sequences in each group and their original counterparts, were randomly selected and split into 96$\times$96 image blocks, which were then converted to the RGB colour space. Block rotation has also been employed for data augmentation to further enhance model generalisation. For each training group, the total number of training image block pairs is approximately 100,000, much more than that of our previous training sets in \cite{j:Zhang9,c:Zhang26}.

The training process was conducted in the Tensorflow environment \cite{w:Tensorflow}, using the following parameters: batch size of 16, learning rate of 0.0001, Adam optimisation \cite{c:Kingma}, weight decay of 0.1 and a total number of 200 training epochs. The number of residual blocks ($N$) is fixed as 16, which is the same as that used in \cite{c:Ledig}. This generates 20 different CNN models ($\mathrm{model}_{c,v,q}$), each for one group. Here $c$ stands for the used codec (HM or VTM), $v$ represents adaptation versions (EBD or SR-EBD), and $q$ is denoted to the QP group. For a specific codec and adaptation version, the CNN model used in evaluation depends on the initial (before applying the offset) base QP ($\mathrm{QP_{base}}$):

\begin{equation}
\left\{
\begin{array}{l r }
\mathrm{model}_{c,v,22}, \  \text{if} & \mathrm{QP_{base}}\leq 24.5 \\
\mathrm{model}_{c,v,27}, \  \text{if} & 24.5<\mathrm{QP_{base}}\leq 29.5 \\
\mathrm{model}_{c,v,32}, \  \text{if} & 29.5<\mathrm{QP_{base}}\leq 34.5 \\
\mathrm{model}_{c,v,37}, \  \text{if} & 34.5<\mathrm{QP_{base}}\leq 39.5 \\
\mathrm{model}_{c,v,42}, \  \text{if}  & \mathrm{QP_{base}}\geq 39.5
\end{array}
\right.
\label{eq:models}
\end{equation}
 
When these CNN models were employed for evaluation, each EBD or SR-EBD down-sampled frame (pre-CNN SR up-sampling has been applied here if SR adaptation is enabled) is firstly segmented into 96$\times$96 overlapping blocks with an overlap size of 4 pixels as CNN input (after converting to the RGB space). The output full SR and EBD blocks were then aggregated back following the same segmentation pattern to form the final reconstructed frame.

\subsection{Quantisation-Resolution Optimisation}

Spatial resolution adaptation does not always lead to coding improvement, being highly dependent on the original resolutions, spatio-temporal characteristics and the host codec. Our previous work \cite{j:Zhang9} employed four video features to predict QP thresholds, beyond which encoding lower resolution content produces higher compression efficiency. In ViSTRA2, an improved machine learning based approach has been developed to make decisions on resolution adaptation, based on a spatial resolution dependent quality metric, SRQM \cite{c:Zhang20}, temporal information (TI) and initial base quantisation parameter. 

SRQM is an efficient, full reference objective video quality metric, which characterises the relationship between variations in spatial resolution and visual quality. It employs wavelet decomposition, subband combination with perceptually inspired weights, and spatio-temporal pooling to estimate the relative quality between the frames of a high resolution reference video, and one that has been spatially adapted through a combination of down- and up-sampling. In this work, SRQM is applied between the uncompressed, original resolution video frames and their re-sampled versions (using Lanczos3 filter for both down- and up- sampling). The temporal information (TI) here is defined as the average absolute frame difference between luma pixels in the current frame and its neighbouring ones (one frame before and another after if available).

Both SRQM and TI values were computed for all frames in the 432 uncompressed sequences in the training dataset at various resolutions from 3840$\times$2160 to 480$\times$270. This database does not include any test sequences used in Section \ref{sec:results}, and was used for training the CNN in Section \ref{sec:training}. The average SRQM and TI values for each sequence, together with tested initial base QP values (22, 27, 32, 37 and 42), were employed as input features to train a fully connected, shadow neural network, which contains a hidden layer and an output layer. 


The target binary output of the network was generated through comparing the rate-distortion performance (PSNR on luma values was used here as a quality metric) of the original anchor codecs (HM and VTM) and that of ViSTRA2 (with the trained CNN models) for all five tested initial base QPs on these training sequences. For a training sequence and $\mathrm{QP_{base}}$, if the rate-PSNR point of ViSTRA2 codec is above the rate-PSNR curve of the corresponding anchor codec, the binary decision (for this sequence and QP) is defined as 1 -- the adaptation should be applied. Otherwise, it is configured to 0. This results in 432$\times$5  binary decisions for each codec. The network was trained offline using the Matlab function \textit{train}, and two sets of network model parameters were then obtained for HEVC HM 16.20 and VVC VTM 4.01.

When these models were employed in ViSTRA2 to predict whether spatial resolution should be enabled, the average feature values used for the QRO are obtained for a number of frames -- one GOP for Random Access mode in HEVC HM and VVC VTM -- in order to maintain the same latency. The decision is applied to the frames assessed. Due to the possibility of content variations (although not very common in the JVET SDR test set), the QRO module may lead to frequent resolution changes during encoding. This should not cause any problem if resolution changing is supported at the frame or GOP level in the encoding loop. It is noted that although an Ad-Hoc Group was established by JVET to investigate resolution adaptivity \cite{s:JVET-O0008}, this feature had not been formally adopted by VVC when this paper was written. Video sequence segmentation is therefore executed in the current version of ViSTRA2, when different resolutions have to be applied during encoding. If a video sequence has to be segmented into more than one segments, their bitstreams are generated sequentially by the host encoder. In order to avoid frequent sequence segmentation, each video segment is constrained here to be longer than 1 second. A flag bit is inserted within the header of the bitstream for each video segment to indicate the decoder resolution change. This solution will be modified when resolution adaptivity feature is integrated into standardised coding algorithms.

\section{Results and Discussion}
\label{sec:results}

\begin{table*}[ht]
\centering
\footnotesize
\caption{BD-rate results of ViSTRA2 when HM 16.20 was employed as host codec.}
\begin{tabular}{l || r | r |r |r | r|r||r|r| r|r|r|r}
\toprule
Metric	& 	\multicolumn{6}{c||}{PSNR} & \multicolumn{6}{c}{VMAF} \\
\midrule Codec&\multicolumn{2}{c|}{ViSTRA2-w/o CNN} & \multicolumn{2}{c|}{EBDA-CNN} & \multicolumn{2}{c||}{ViSTRA2-CNN} & \multicolumn{2}{c|}{ViSTRA2-w/o CNN} & \multicolumn{2}{c|}{EBDA-CNN} & \multicolumn{2}{c}{ViSTRA2-CNN}\\ 
\midrule	  
Class-Sequence &   H-QPs 		&    L-QPs	&   H-QPs 		&    L-QPs	&   H-QPs 		&    L-QPs	&   H-QPs 		&    L-QPs	&   H-QPs 		&    L-QPs	&   H-QPs 		&    L-QPs	 \\
\midrule \midrule 
A-Campfire&-28.0\% & +2.0\% &-22.0\% & -16.7\% &-35.5\% & -19.9\% &-28.9\% & -19.6\% &-24.2\% & -25.5\% &-42.2\% & -41.2\% \\ 
A-FoodMarket4&-12.2\% & -3.5\% &-7.9\% & -7.2\% &-18.0\% & -11.8\% &-14.4\% & -12.5\% &-12.2\% & -12.2\% &-24.5\% & -20.9\% \\ 
A-Tango2&-11.9\% & -6.8\% &-8.6\% & -9.7\% &-17.4\% & -15.2\% &-11.9\% & -10.0\% &-12.5\% & -15.4\% &-22.3\% & -20.7\% \\ 
A-CatRobot1&-4.4\% & -0.3\% &-11.9\% & -12.3\% &-15.1\% & -13.1\% &-6.3\% & -3.7\% &-18.4\% & -22.2\% &-23.1\% & -23.7\% \\ 
A-DaylightRoad2&+1.1\% & +2.4\% &-10.6\% & -14.3\% &-11.2\% & -13.7\% &-0.5\% & -2.1\% &-17.5\% & -25.3\% &-21.7\% & -26.5\% \\ 
A-ParkRunning3&-26.5\% & -17.4\% &-18.5\% & -13.7\% &-31.2\% & -25.2\% &-29.7\% & -27.2\% &-21.7\% & -20.4\% &-36.7\% & -34.9\% \\ 
\midrule \textbf{Class A}& \textbf{-13.7\% }& \textbf{-4.0\% }& \textbf{-13.2\% }& \textbf{-12.3\% }& \textbf{-21.4\% }& \textbf{-16.5\% }& \textbf{-15.3\% }& \textbf{-12.5\% }& \textbf{-17.8\% }& \textbf{-20.2\% }& \textbf{-28.4\% }& \textbf{-28.0\% }\\ 
 \midrule B-BQTerrace&-0.3\% & -0.2\% &-11.0\% & -9.8\% &-11.0\% & -9.8\% &-1.1\% & -2.8\% &-21.7\% & -28.2\% &-21.7\% & -28.2\% \\ 
B-BasketballDrive&-1.3\% & -1.7\% &-11.0\% & -10.5\% &-10.0\% & -10.5\% &-2.9\% & -4.9\% &-11.7\% & -14.2\% &-12.6\% & -14.2\% \\ 
B-Cactus&-0.4\% & -0.3\% &-9.5\% & -9.8\% &-9.5\% & -9.8\% &-3.7\% & -4.9\% &-15.2\% & -18.6\% &-16.8\% & -18.6\% \\ 
B-MarketPlace&-0.7\% & +2.3\% &-4.8\% & -4.5\% &-6.7\% & -4.7\% &-3.3\% & -0.8\% &-12.4\% & -13.7\% &-18.0\% & -15.9\% \\ 
B-RitualDance&-1.8\% & +1.2\% &-8.0\% & -7.0\% &-9.8\% & -7.2\% &-2.0\% & -1.2\% &-12.7\% & -13.6\% &-16.1\% & -14.7\% \\ 
\midrule \textbf{Class B}& \textbf{-0.9\% }& \textbf{+0.3\% }& \textbf{-8.9\% }& \textbf{-8.3\% }& \textbf{-9.4\% }& \textbf{-8.4\% }& \textbf{-2.6\% }& \textbf{-2.9\% }& \textbf{-14.7\% }& \textbf{-17.7\% }& \textbf{-17.0\% }& \textbf{-18.3\% }\\ 
 \midrule C-BQMall&+0.1\% & +1.6\% &-9.5\% & -8.7\% &-9.5\% & -8.7\% &-1.0\% & -0.5\% &-13.1\% & -13.3\% &-13.1\% & -13.3\% \\ 
C-BasketballDrill&-2.1\% & +1.2\% &-11.9\% & -8.8\% &-11.9\% & -8.8\% &-5.5\% & -4.4\% &-15.6\% & -13.8\% &-15.6\% & -13.8\% \\ 
C-PartyScene&+0.0\% & +1.6\% &-8.0\% & -7.4\% &-8.0\% & -7.4\% &-1.1\% & -1.1\% &-11.2\% & -12.4\% &-11.2\% & -12.4\% \\ 
C-RaceHorses&-3.5\% & -2.0\% &-10.5\% & -8.4\% &-10.5\% & -8.4\% &-4.6\% & -5.4\% &-12.7\% & -13.4\% &-12.7\% & -13.4\% \\ 
\midrule \textbf{Class C}& \textbf{-1.4\% }& \textbf{+0.6\% }& \textbf{-10.0\% }& \textbf{-8.3\% }& \textbf{-10.0\% }& \textbf{-8.3\% }& \textbf{-3.0\% }& \textbf{-2.9\% }& \textbf{-13.1\% }& \textbf{-13.2\% }& \textbf{-13.1\% }& \textbf{-13.2\% }\\ 
 \midrule D-BQSquare&+0.8\% & +2.1\% &-15.6\% & -15.8\% &-15.6\% & -15.8\% &-1.3\% & -2.3\% &-18.1\% & -22.7\% &-18.1\% & -22.7\% \\ 
D-BasketballPass&-3.2\% & -0.8\% &-11.7\% & -9.8\% &-11.7\% & -9.8\% &-4.7\% & -4.3\% &-13.1\% & -12.3\% &-13.1\% & -12.3\% \\ 
D-BlowingBubbles&-0.4\% & +1.3\% &-8.0\% & -7.1\% &-8.0\% & -7.1\% &-2.4\% & -1.3\% &-12.5\% & -12.0\% &-12.5\% & -12.0\% \\ 
D-RaceHorses&-3.0\% & -1.2\% &-11.2\% & -9.3\% &-11.2\% & -9.3\% &-5.2\% & -5.4\% &-14.0\% & -14.2\% &-14.0\% & -14.2\% \\ 
\midrule \textbf{Class D}& \textbf{-1.5\% }& \textbf{+0.3\% }& \textbf{-11.6\% }& \textbf{-10.5\% }& \textbf{-11.6\% }& \textbf{-10.5\% }& \textbf{-3.4\% }& \textbf{-3.3\% }& \textbf{-14.4\% }& \textbf{-15.3\% }& \textbf{-14.4\% }& \textbf{-15.3\% }\\ 
 \midrule \midrule \textbf{Overall}& \textbf{-5.2\% }& \textbf{-1.0\% }& \textbf{-11.1\% }& \textbf{-10.0\% }& \textbf{-13.8\% }& \textbf{-11.4\% }& \textbf{-6.9\% }& \textbf{-6.0\% }& \textbf{-15.3\% }& \textbf{-17.0\% }& \textbf{-19.3\% }& \textbf{-19.7\% }\\ 
\cmidrule{2-13}
&\multicolumn{2}{c|}{\textbf{-3.1\%}}	& \multicolumn{2}{c|}{\textbf{-10.6\%}} & \multicolumn{2}{c||}{\textbf{-12.6\%}} &\multicolumn{2}{c|}{\textbf{-6.5\%}}	& \multicolumn{2}{c|}{\textbf{-16.2\%}} & \multicolumn{2}{c}{\textbf{-19.5\%}}\\ 
\bottomrule	
\end{tabular}
\label{tab:results1}
\end{table*}

\begin{table*}[ht]
\centering
\footnotesize
\caption{BD-rate results of ViSTRA2 when VTM 4.01 was employed as host codec.}
\begin{tabular}{l || r | r |r |r | r|r||r|r| r|r|r|r}
\toprule
Metric	& 	\multicolumn{6}{c||}{PSNR} & \multicolumn{6}{c}{VMAF} \\
\midrule Codec&\multicolumn{2}{c|}{ViSTRA2-w/o CNN} & \multicolumn{2}{c|}{EBDA-CNN} & \multicolumn{2}{c||}{ViSTRA2-CNN} & \multicolumn{2}{c|}{ViSTRA2-w/o CNN} & \multicolumn{2}{c|}{EBDA-CNN} & \multicolumn{2}{c}{ViSTRA2-CNN}\\ 
\midrule	  
Class-Sequence &   H-QPs 		&    L-QPs	&   H-QPs 		&    L-QPs	&   H-QPs 		&    L-QPs	&   H-QPs 		&    L-QPs	&   H-QPs 		&    L-QPs	&   H-QPs 		&    L-QPs	 \\
\midrule \midrule 
A-Campfire&-20.2\% & -9.0\% &-11.4\% & -11.3\% &-30.3\% & -16.8\% &-24.6\% & -21.3\% &-16.7\% & -20.9\% &-38.3\% & -36.5\% \\ 
A-FoodMarket4&-3.9\% & +2.0\% &-2.5\% & -0.6\% &-7.3\% & -1.9\% &-11.0\% & -6.2\% &-5.6\% & -3.4\% &-14.7\% & -9.6\% \\ 
A-Tango2&-4.6\% & +0.1\% &-3.0\% & -1.9\% &-7.9\% & -3.9\% &-9.4\% & -5.1\% &-5.5\% & -5.2\% &-12.3\% & -8.4\% \\ 
A-CatRobot1&+4.0\% & +4.7\% &-5.3\% & -3.6\% &-4.8\% & -3.3\% &+0.4\% & -0.0\% &-7.3\% & -7.9\% &-9.0\% & -8.0\% \\ 
A-DaylightRoad2&+1.9\% & +2.9\% &-4.1\% & -5.8\% &-5.0\% & -5.8\% &-0.3\% & -1.5\% &-7.1\% & -11.1\% &-9.1\% & -11.1\% \\ 
A-ParkRunning3&-19.1\% & -17.0\% &-15.7\% & -17.0\% &-24.0\% & -25.0\% &-22.1\% & -19.7\% &-16.7\% & -16.3\% &-26.6\% & -25.7\% \\ 
\midrule \textbf{Class A}& \textbf{-7.0\% }& \textbf{-2.7\% }& \textbf{-7.0\% }& \textbf{-6.7\% }& \textbf{-13.2\% }& \textbf{-9.5\% }& \textbf{-11.2\% }& \textbf{-9.0\% }& \textbf{-9.8\% }& \textbf{-10.8\% }& \textbf{-18.3\% }& \textbf{-16.5\% }\\ 
 \midrule B-BQTerrace&+1.6\% & +4.5\% &-1.7\% & +0.1\% &-1.7\% & +0.1\% &+0.3\% & +0.8\% &-1.5\% & -3.8\% &-1.5\% & -3.8\% \\ 
B-BasketballDrive&-0.8\% & +0.3\% &-4.0\% & -3.9\% &-4.0\% & -3.9\% &-2.6\% & -3.5\% &-3.4\% & -5.1\% &-3.4\% & -5.1\% \\ 
B-Cactus&-1.5\% & -0.4\% &-4.4\% & -4.0\% &-4.4\% & -4.0\% &-3.7\% & -4.6\% &-7.3\% & -8.7\% &-7.3\% & -8.7\% \\ 
B-MarketPlace&+7.1\% & +7.3\% &+4.4\% & +4.2\% &+4.4\% & +4.2\% &+5.3\% & +4.1\% &-0.7\% & -1.8\% &-0.7\% & -1.8\% \\ 
B-RitualDance&+0.4\% & +2.0\% &-3.2\% & -2.2\% &-3.2\% & -2.2\% &-1.3\% & -1.9\% &-6.7\% & -6.7\% &-6.7\% & -6.7\% \\ 
\midrule \textbf{Class B}& \textbf{+1.4\% }& \textbf{+2.7\% }& \textbf{-1.8\% }& \textbf{-1.2\% }& \textbf{-1.8\% }& \textbf{-1.2\% }& \textbf{-0.4\% }& \textbf{-1.0\% }& \textbf{-3.9\% }& \textbf{-5.2\% }& \textbf{-3.9\% }& \textbf{-5.2\% }\\ 
 \midrule C-BQMall&+3.6\% & +4.9\% &-2.2\% & -2.1\% &-2.2\% & -2.1\% &+2.3\% & +2.1\% &-2.5\% & -2.8\% &-2.5\% & -2.8\% \\ 
C-BasketballDrill&+3.3\% & +7.2\% &-1.8\% & +0.7\% &-1.8\% & +0.7\% &-0.7\% & +0.3\% &-3.3\% & -2.7\% &-3.3\% & -2.7\% \\ 
C-PartyScene&+0.9\% & +2.6\% &-4.3\% & -3.2\% &-4.3\% & -3.2\% &-0.3\% & +0.4\% &-4.8\% & -3.7\% &-4.8\% & -3.7\% \\ 
C-RaceHorses&-1.4\% & -0.6\% &-4.0\% & -3.3\% &-4.0\% & -3.3\% &-3.4\% & -3.9\% &-6.3\% & -6.9\% &-6.3\% & -6.9\% \\ 
\midrule \textbf{Class C}& \textbf{+1.6\% }& \textbf{+3.5\% }& \textbf{-3.1\% }& \textbf{-2.0\% }& \textbf{-3.1\% }& \textbf{-2.0\% }& \textbf{-0.5\% }& \textbf{-0.2\% }& \textbf{-4.2\% }& \textbf{-4.0\% }& \textbf{-4.2\% }& \textbf{-4.0\% }\\ 
 \midrule D-BQSquare&+5.7\% & +11.1\% &-5.7\% & -1.8\% &-5.7\% & -1.8\% &+3.9\% & +12.1\% &-1.8\% & +6.4\% &-1.8\% & +6.4\% \\ 
D-BasketballPass&+0.1\% & +1.9\% &-5.6\% & -4.6\% &-5.6\% & -4.6\% &-0.6\% & -0.6\% &-6.2\% & -5.4\% &-6.2\% & -5.4\% \\ 
D-BlowingBubbles&+1.1\% & +2.6\% &-3.5\% & -2.6\% &-3.5\% & -2.6\% &-0.5\% & +0.0\% &-5.4\% & -4.0\% &-5.4\% & -4.0\% \\ 
D-RaceHorses&-1.9\% & -0.5\% &-6.5\% & -5.6\% &-6.5\% & -5.6\% &-5.9\% & -6.0\% &-10.3\% & -10.4\% &-10.3\% & -10.4\% \\ 
\midrule \textbf{Class D}& \textbf{+1.3\% }& \textbf{+3.8\% }& \textbf{-5.3\% }& \textbf{-3.7\% }& \textbf{-5.3\% }& \textbf{-3.7\% }& \textbf{-0.8\% }& \textbf{+1.4\% }& \textbf{-5.9\% }& \textbf{-3.4\% }& \textbf{-5.9\% }& \textbf{-3.4\% }\\ 
 \midrule \midrule \textbf{Overall}& \textbf{-1.2\% }& \textbf{+1.4\% }& \textbf{-4.5\% }& \textbf{-3.6\% }& \textbf{-6.4\% }& \textbf{-4.5\% }& \textbf{-3.9\% }& \textbf{-2.9\% }& \textbf{-6.3\% }& \textbf{-6.3\% }& \textbf{-9.0\% }& \textbf{-8.2\% }\\ 
\cmidrule{2-13}
&\multicolumn{2}{c|}{\textbf{1.3\%}}	& \multicolumn{2}{c|}{\textbf{-4.1\%}} & \multicolumn{2}{c||}{\textbf{-5.5\%}} &\multicolumn{2}{c|}{\textbf{-3.4\%}}	& \multicolumn{2}{c|}{\textbf{-6.3\%}} & \multicolumn{2}{c}{\textbf{-8.6\%}}\\  
\bottomrule	
\end{tabular}
\label{tab:results2}
\end{table*}

The ViSTRA2 coding framework has been integrated into the HEVC (HM 16.20) and VVC (VTM 4.01) reference software, and has been fully tested under JVET CTC \cite{s:JVETCTC} using the Random Access configuration (Main10 profile). In order to evaluate the proposed approach across a wider bitrate range, the initial base QPs tested include 22, 27, 32, 37 and 42. The SDR (standard dynamic range) video classes A1, A2, B, C and D from JVET CTC were employed here as test content, none of which were utilised in the training of CNN and QRO modules in Section \ref{sec:algorithm}.

The rate quality performance of the integrated ViSTRA2 has been compared to the corresponding reference codecs. Bj{\o}ntegaard delta measurements (BD) \cite{r:Bjontegaard} results were generated for both low QP (22, 27, 32 and 37) and high QP (27, 32, 37 and 42) values over all frames using two quality metrics (on luma values only), Peak Signal-to-Noise Ratio (PSNR) and Video Multimethod Assessment Fusion (VMAF, version 0.6.1) \cite{w:VMAF}. PSNR is the most commonly used image quality metric for evaluating video compression performance, while the recently developed VMAF is a machine learning based assessment method, combining the Visual Information Fidelity measures (VIF) \cite{j:Sheikh}, Detail Loss Metric (DLM) \cite{j:Li2} and temporal frame difference \cite{w:VMAF} together using a Support Vector Machine (SVM) regressor \cite{j:Cortes}. As the content generated by ViSTRA2 contains both compression distortions and resolution re-sampled artefacts, VMAF, which is one of few quality metrics being trained and evaluated on compressed and resolution adapted content, is expected to provide more reliable assessment results (better correlation with subjective opinions) than PSNR. 

In order to further benchmark the contribution of CNN reconstruction, results through up-sampling using Lanczos3 \cite{b:Turkowski} (for SR) and simple bitwise left shifting (for EBD) were also produced (denoted ViSTRA2-w/o CNN), based on the same QRO decisions (as CNN-based ViSTRA2). Moreover, the results generated by CNN-based EBD adaptation only (denoted EBDA-CNN) were presented as well, providing another benchmark for the full ViSTRA2.

Finally, the computational complexity of ViSTRA2 was calculated and normalised to the corresponding anchor codecs. The encoding process was executed on a shared cluster, BlueCrystal Phase3 \cite{w:BC3} based in the University of Bristol, in which each CPU node has 16 $\times$2.6 GHz SandyBridge cores and 64GB RAM. The decoding was run on the GPU nodes of the BlueCrystal Phase 4 \cite{w:BC4}, each of which has 14 core 2.4 GHz Intel E5-2680 v4 (Broadwell) CPUs, 138GB of RAM and two NVIDIA P100 graphic cards. 

\subsection{Compression results}

\begin{figure*}[ht]
\centering
\footnotesize
\centering
\begin{minipage}[b]{0.245\linewidth}
\centering
\centerline{\includegraphics[width=1.1\linewidth]{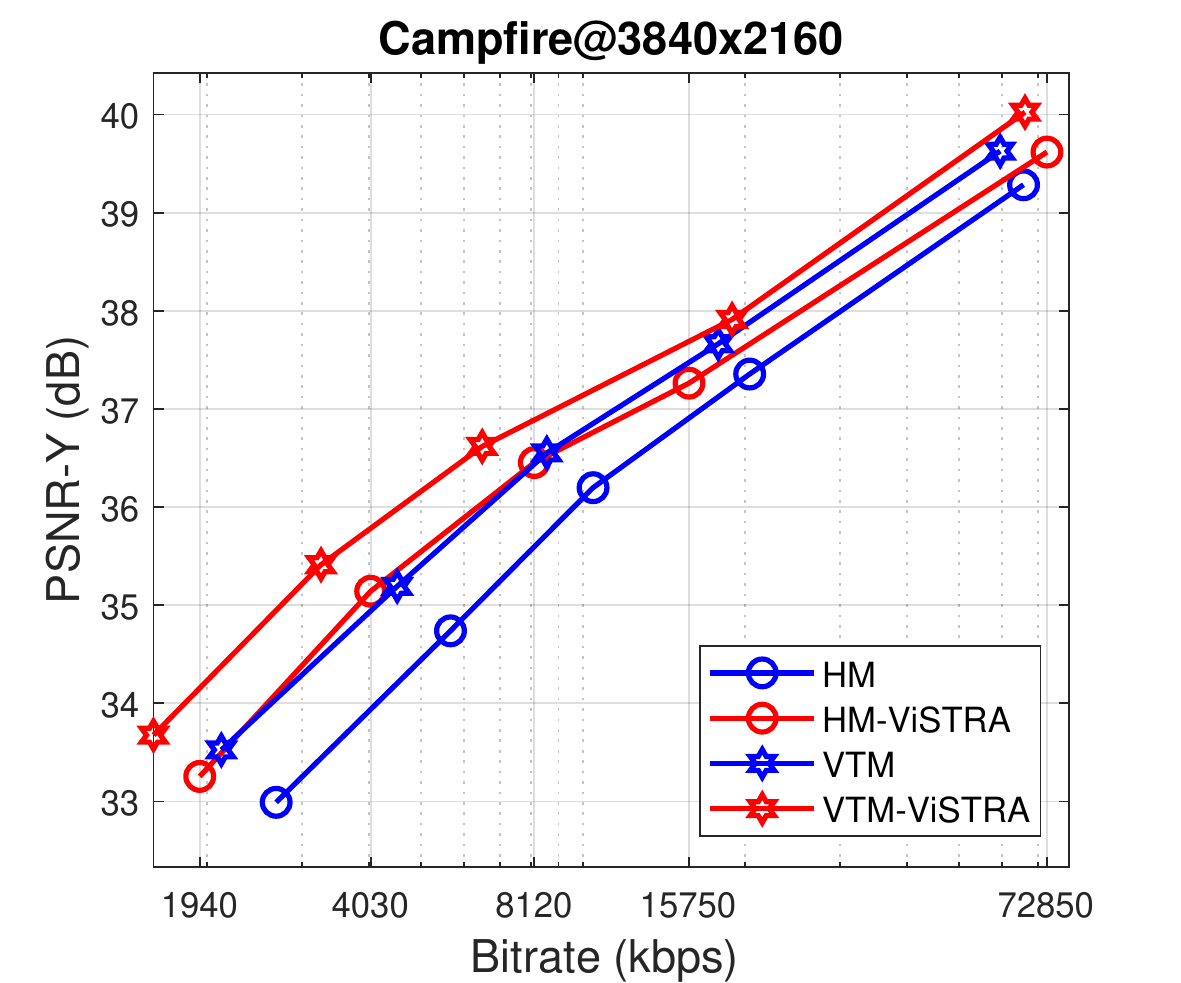}}
(a) Campfile-PSNR
\end{minipage}
\begin{minipage}[b]{0.245\linewidth}
\centering
\centerline{\includegraphics[width=1.1\linewidth]{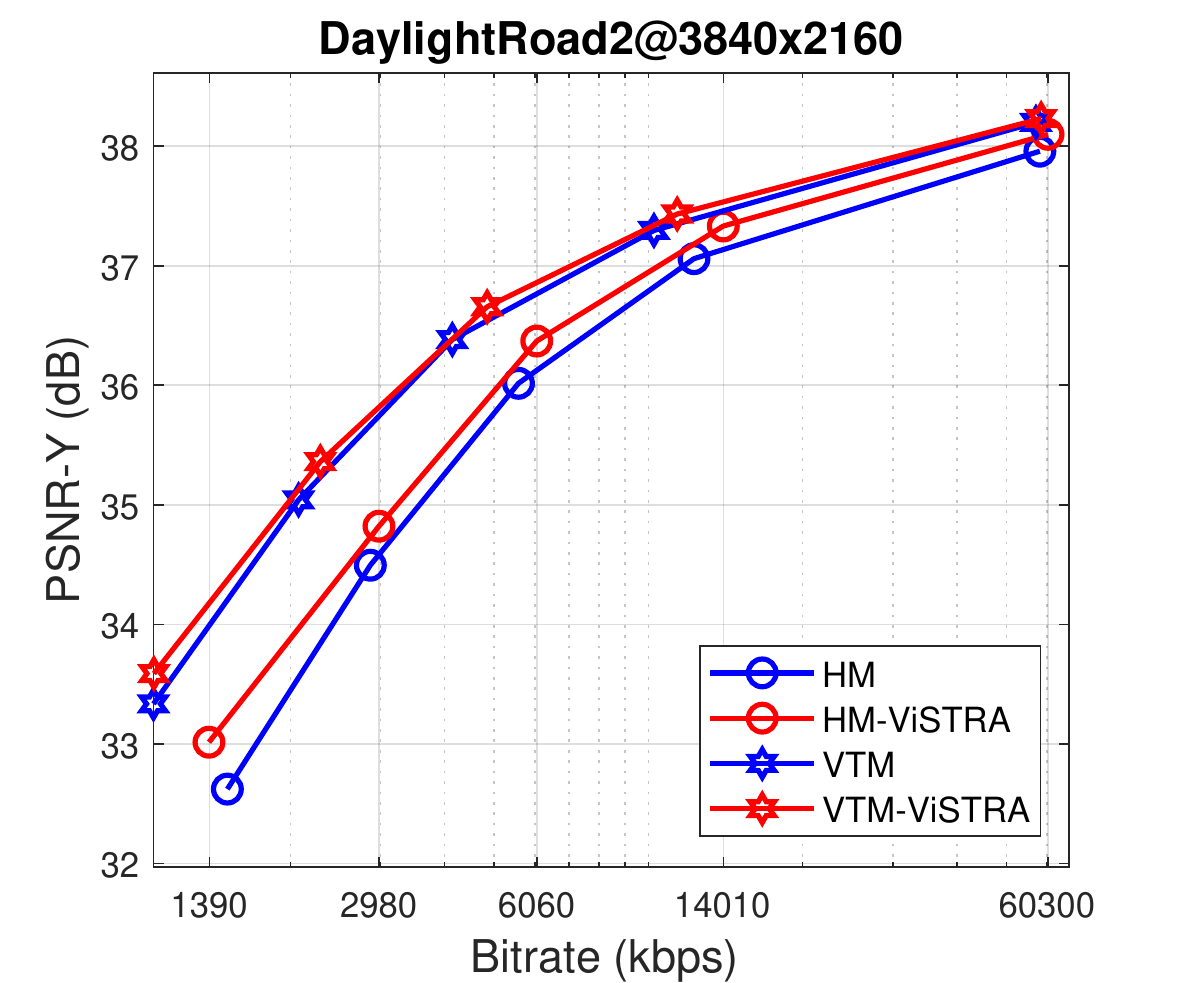}}
(b) DaylightRoad2-PSNR
\end{minipage}
\begin{minipage}[b]{0.245\linewidth}
\centering
\centerline{\includegraphics[width=1.1\linewidth]{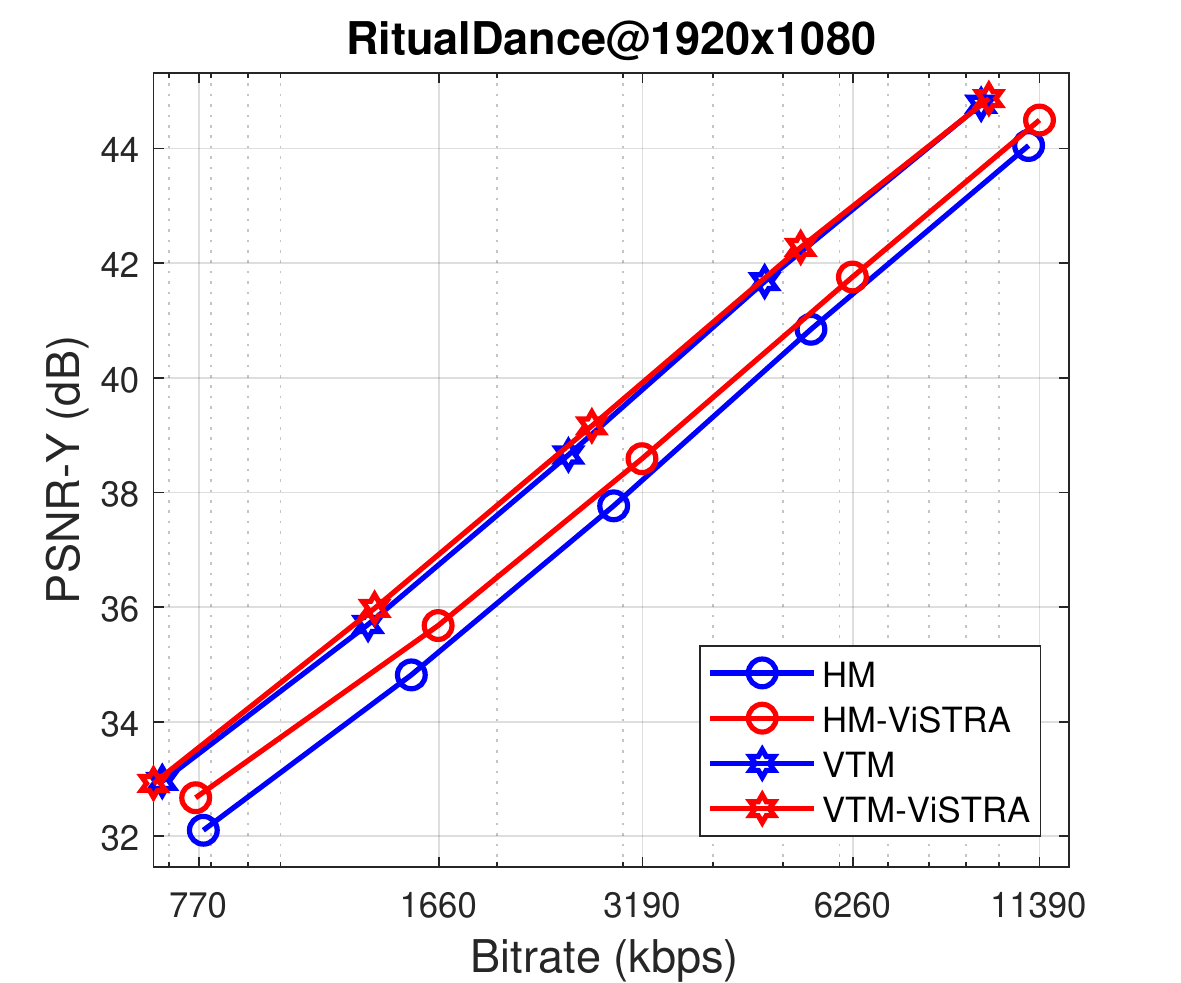}}
(c) RitualDance-PSNR
\end{minipage}
\begin{minipage}[b]{0.245\linewidth}
\centering
\centerline{\includegraphics[width=1.1\linewidth]{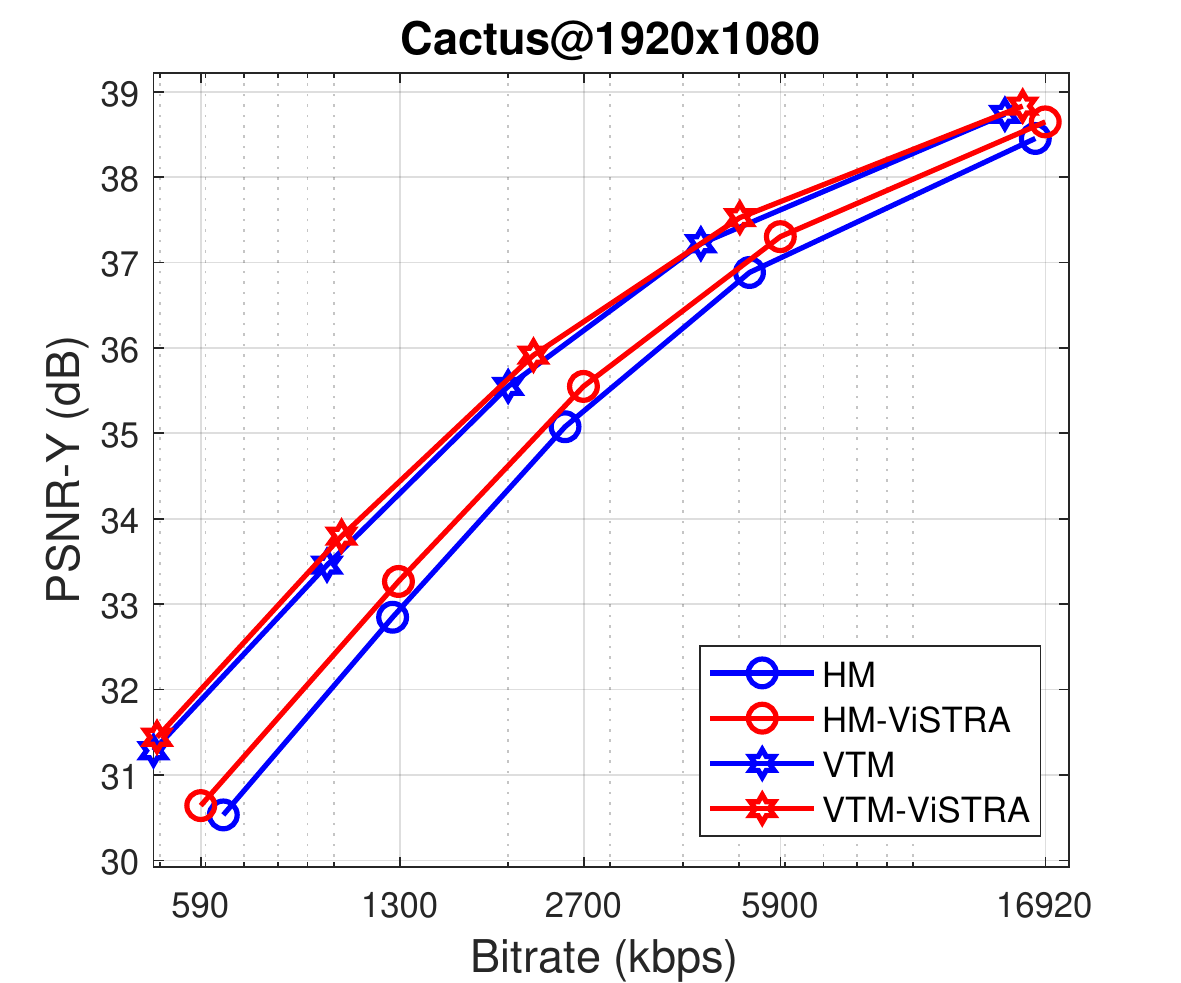}}
(d) Cactus-PSNR
\end{minipage}
\begin{minipage}[b]{0.245\linewidth}
\centering
\centerline{\includegraphics[width=1.1\linewidth]{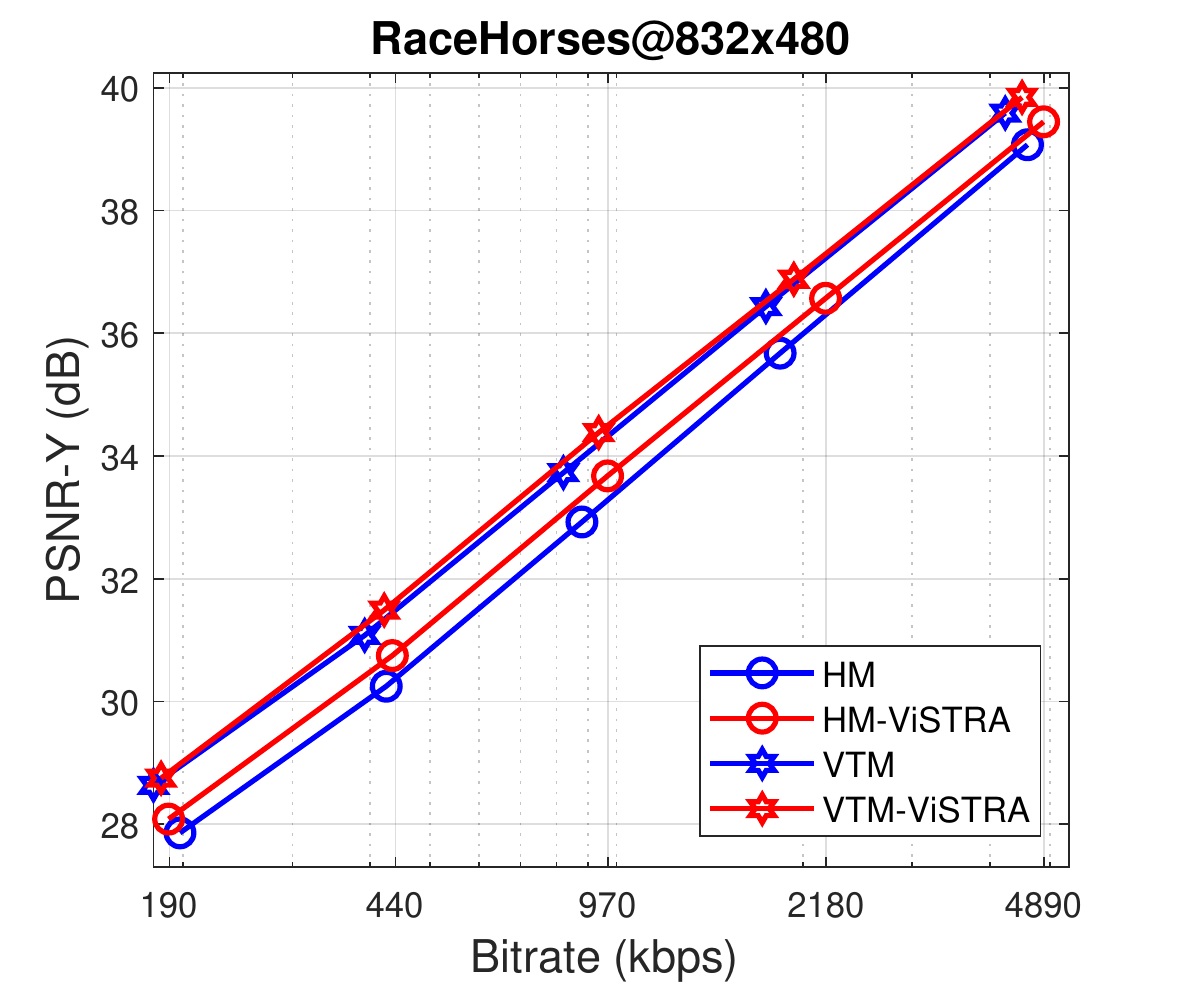}}
(e) RaceHorses-PSNR
\end{minipage}
\begin{minipage}[b]{0.245\linewidth}
\centering
\centerline{\includegraphics[width=1.1\linewidth]{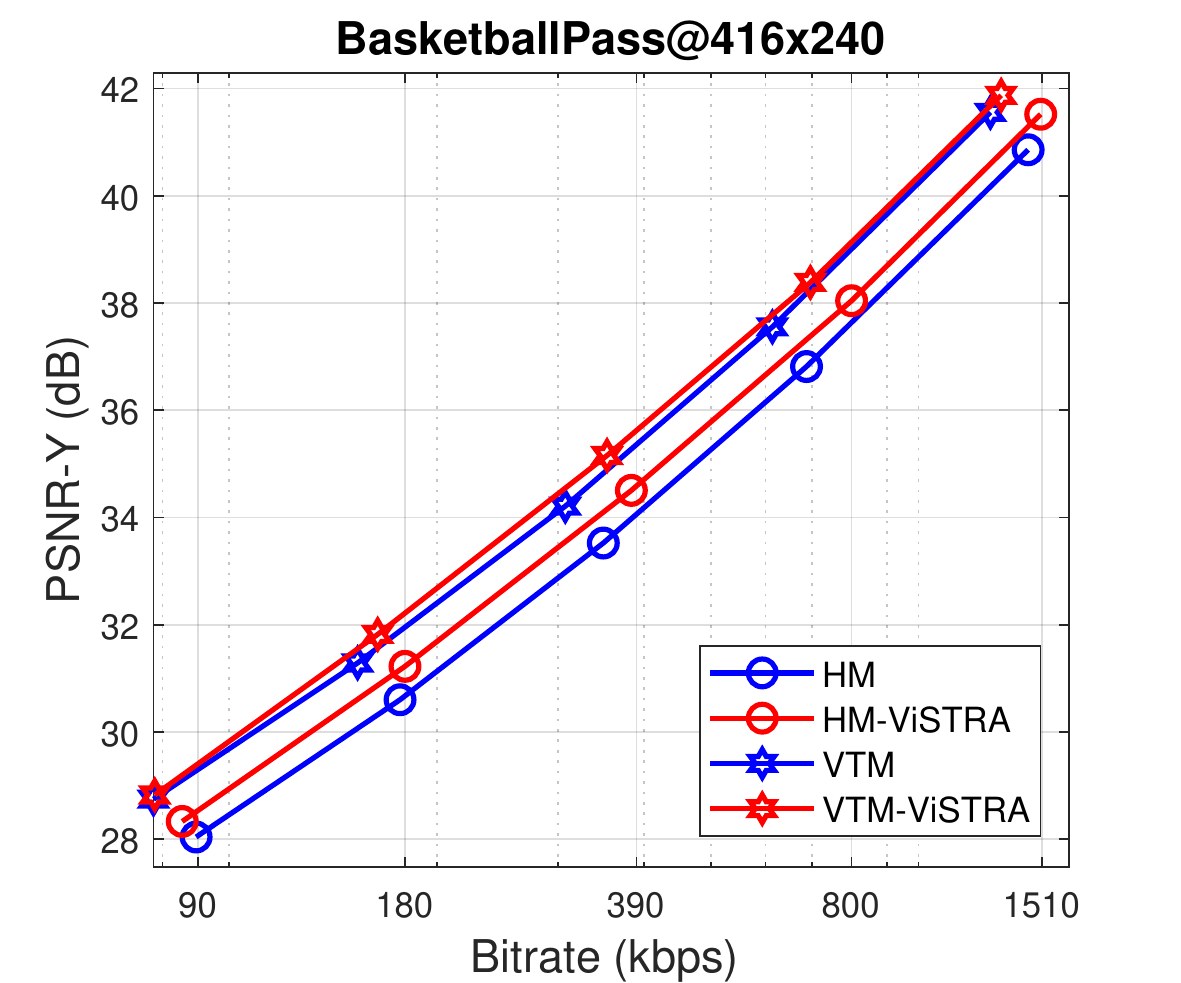}}
(f) BasketballPass-PSNR
\end{minipage}
\begin{minipage}[b]{0.245\linewidth}
\centering
\centerline{\includegraphics[width=1.1\linewidth]{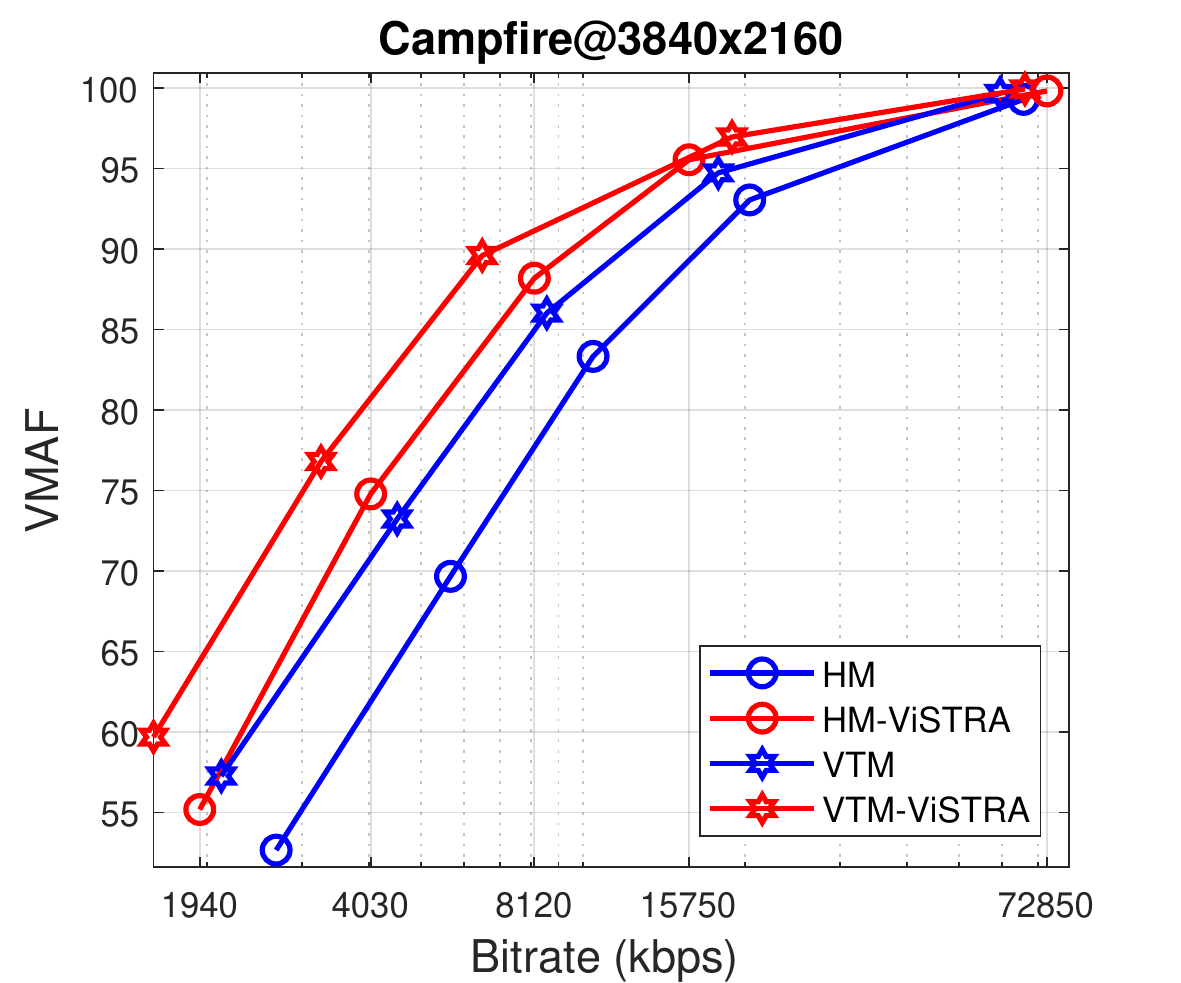}}
(g) Campfile-VMAF
\end{minipage}
\begin{minipage}[b]{0.245\linewidth}
\centering
\centerline{\includegraphics[width=1.1\linewidth]{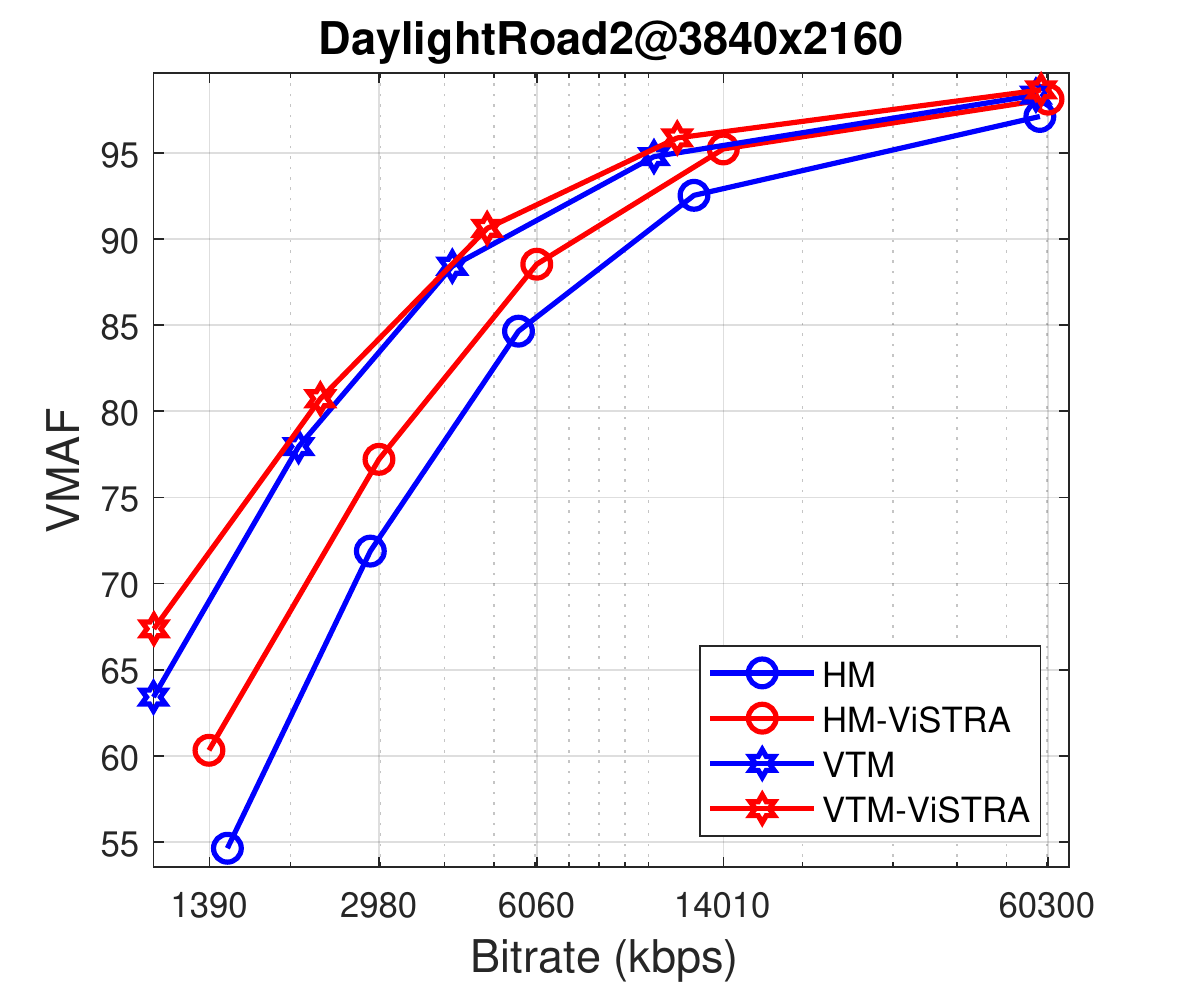}}
(h) DaylightRoad2-VMAF
\end{minipage}
\begin{minipage}[b]{0.245\linewidth}
\centering
\centerline{\includegraphics[width=1.1\linewidth]{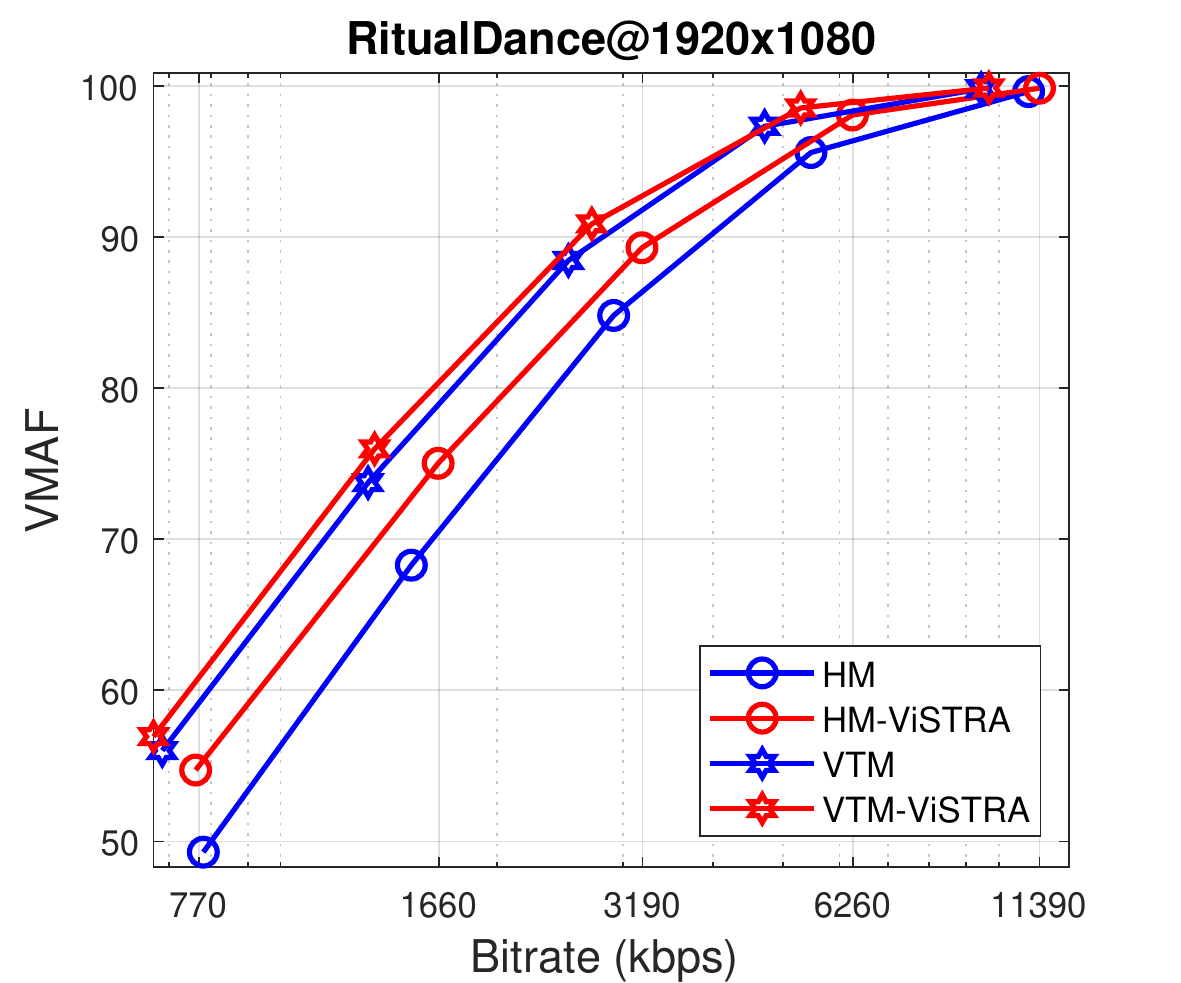}}
(i) RitualDance-VMAF
\end{minipage}
\begin{minipage}[b]{0.245\linewidth}
\centering
\centerline{\includegraphics[width=1.1\linewidth]{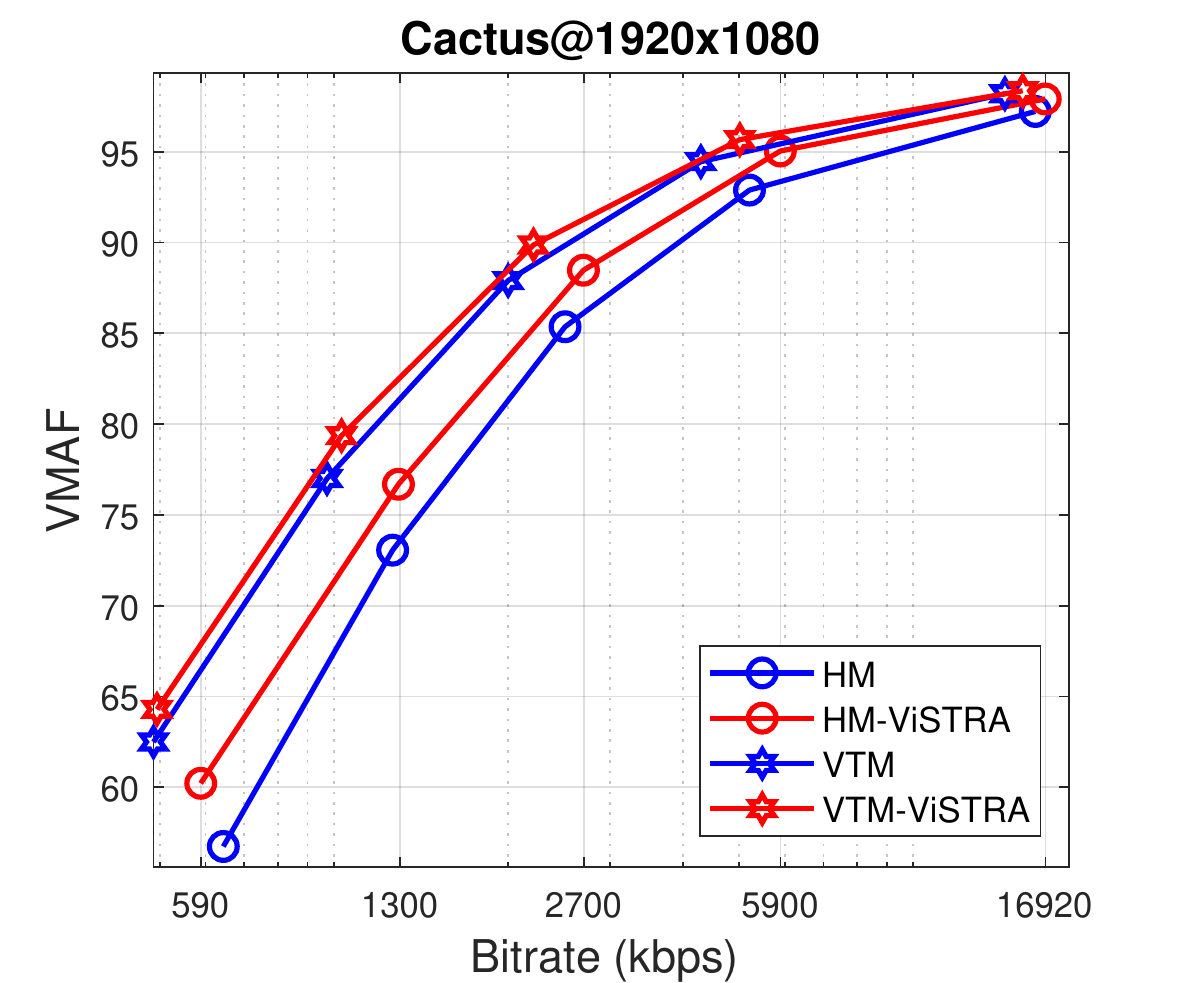}}
(j) Cactus-VMAF
\end{minipage}
\begin{minipage}[b]{0.245\linewidth}
\centering
\centerline{\includegraphics[width=1.1\linewidth]{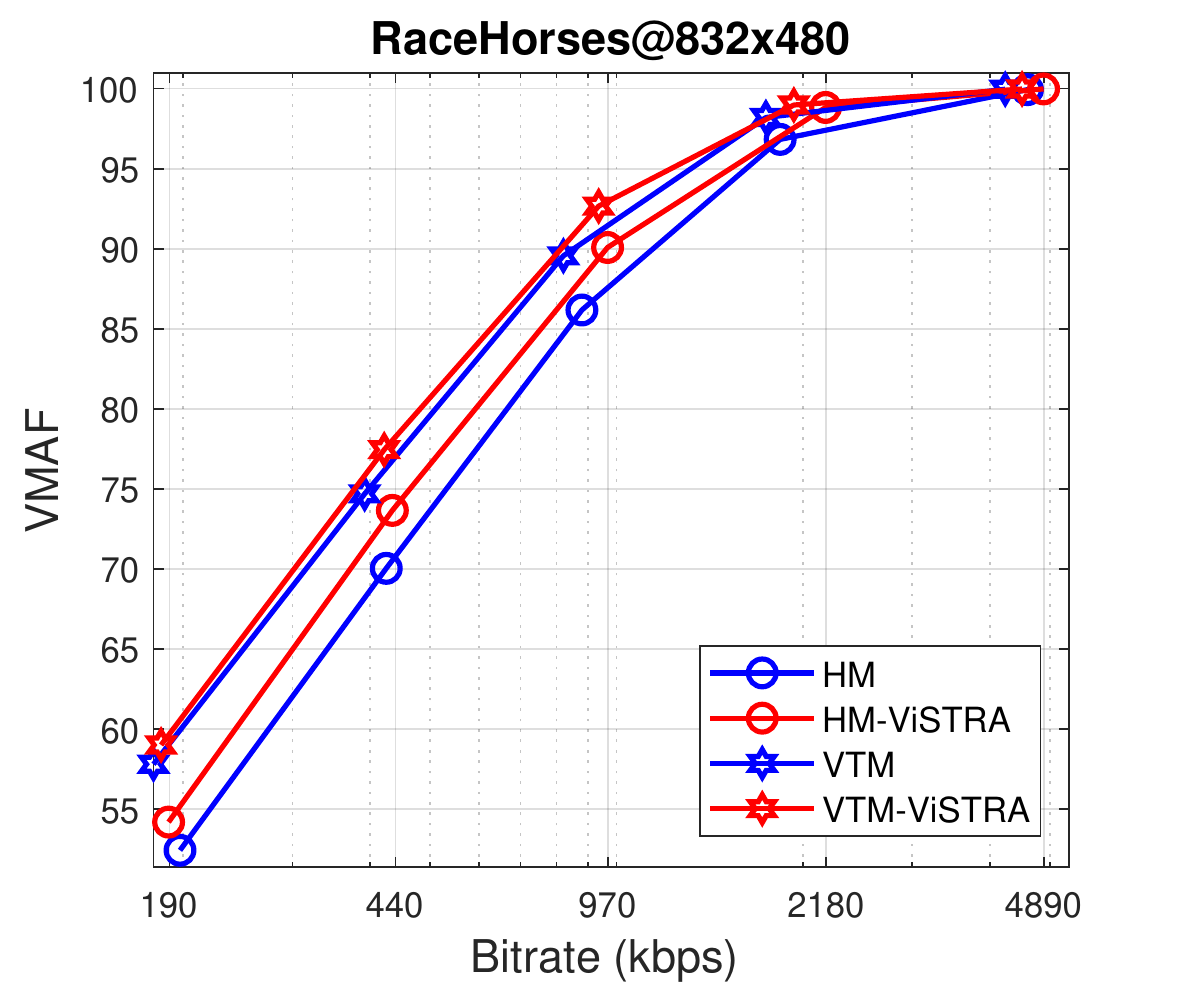}}
(k) RaceHorses-VMAF
\end{minipage}
\begin{minipage}[b]{0.245\linewidth}
\centering
\centerline{\includegraphics[width=1.1\linewidth]{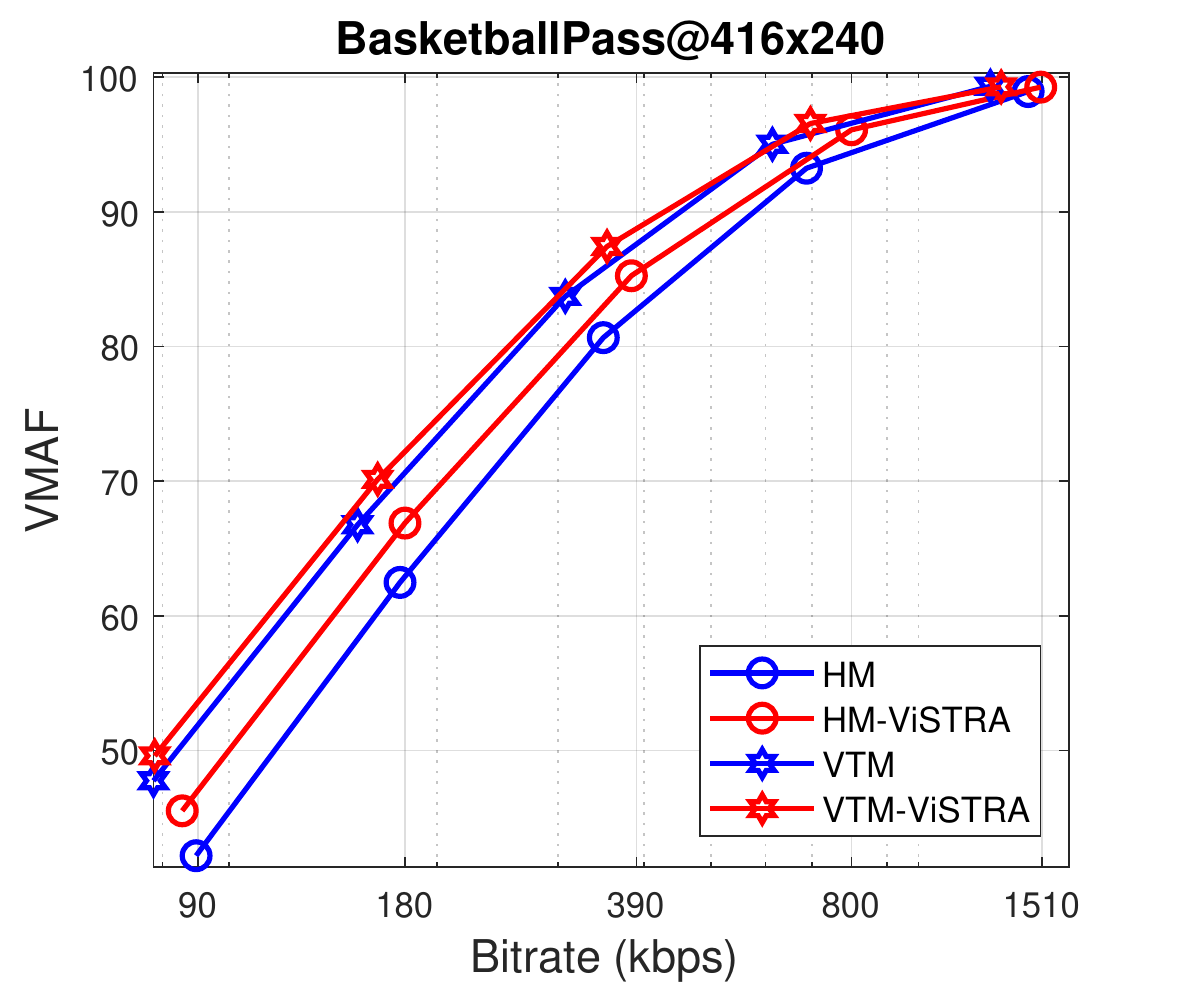}}
(l) BasketballPass-VMAF
\end{minipage}
\caption{Rate-PSNR and Rate-VMAF curves for six selected test sequences.}
\label{fig:curves}
\end{figure*}

Tables \ref{tab:results1} and \ref{tab:results2} summarise the compression results of ViSTRA2 for JVET CTC test sequences when integrated with HM 16.20 and VTM 4.01, using PSNR and VMAF as quality metrics. The rate-quality curves for selected sequences are also shown in Fig. \ref{fig:curves}. It can be observed that ViSTRA2 provides consistent bitrate savings for both host codecs based on PSNR, with average BD-rate values of -12.6\% for HM and -5.5\% for VTM. The coding gains are greater when perceptual quality metric VMAF is employed for quality assessment, and the BD-rate savings (based on VMAF) are -19.5\% and -8.6\% for HM and VTM respectively. The coding improvement from ViSTRA2 for VVC VTM is relatively lower (although consistent) than that for HEVC HM. This is likely to be due to the significant enhancements already achieved by VTM over HM. 

Comparing to ViSTRA2-w/o CNN, ViSTRA2 is superior in terms of overall coding efficiency for all test sequences at all resolutions. This improvement is even higher when VMAF is used to calculate the rate quality performance. It is also noted that the overall compression performance of EBDA-CNN (without SR adaptation) is worse than that of the full ViSTRA2, especially for UHD content, although the former has already been improved from our previous work \cite{c:Zhang26} due to the much larger training set used. Comparing to the compression results presented in \cite{j:Zhang9,s:Zhang1}, where the improvement is only evident on high resolution and low QP cases, ViSTRA2 offers enhanced coding performance across a wider QP and resolution range. This is due to the more sophisticated network architecture and more diverse training data employed.

Finally, it can also be observed that ViSTRA2 performs best on UHD sequences (Class A), where both SR and EBD adaptation were enabled for most tested QP values. For lower resolutions, SR adaptation was less frequently activated, and the coding performance of ViSTRA2-CNN is therefore very close (or identical) to that of EBDA-CNN. Between the two tested QP ranges, ViSTRA2 offers an improved performance for higher QP cases than in low QP scenarios. This becomes more evident on UHD content (Class A), when spatial resolution adaptation was more frequently employed.

\subsection{Complexity figures}
\begin{table}[htbp]
\centering
\footnotesize
\caption{Computational Complexity of  ViSTRA2.}
\begin{tabular}{l | c | c | c | c}
\toprule
\multirow{2}{*}{Host codec}	& \multicolumn{2}{c}{HM 16.20} & \multicolumn{2}{c}{VTM 4.01}	\\ 
\cmidrule{2-5}
	  & Encoder & Decoder & Encoder & Decoder\\
		\midrule
\midrule Class A &   57\% & 2,168\% & 91\% & 1,842\%\\
\midrule Class B &   78\% & 4,153\% & 102\% & 3,690\%\\
\midrule Class C &   102\% & 9,273\% & 99\% & 7,679\%\\
\midrule Class D &   102\% & 12,785\% & 103\% & 11,662\%\\
\midrule Average &   80\% & 6,421\% & 98\% & 5,625\%\\
\bottomrule	
\end{tabular}
\label{tab:complexity}
\end{table}

The complexity figures of ViSTRA2 are presented in Table \ref{tab:complexity}. It is important to note the machine learning architecture employed in ViSTRA2 takes account of the simple filter-based re-sampling process employed in the encoder. This means that, although we benefit from the reconstruction power of a deep network, the encoding process complexity remains similar to the reference encoder. In fact, the average encoding time of ViSTRA2 is shorter (if SR adaptation is enabled) or equivalent to that of the corresponding anchor codecs. This is due to the encoder processing down-sampled frames, but also because we do not employ a neural network at the encoder. During decoding process, the use of CNN-based up-sampling has significantly increased the execution time, on average 64 times that of HM and 56 times that of VTM.

\section{Conclusion}
\label{sec:conclusion}

A new video coding framework (ViSTRA2) has been presented using CNN-based spatial resolution and effective bit depth re-sampling. This approach adaptively reduces the spatial resolution and effective bit depth of input video content for encoding, and employs a deep CNN at the decoder to reconstruct its original format. ViSTRA2 has been integrated with HEVC HM 16.20 and VVC VTM 4.01 reference software, showing consistent coding gains for test sequences at various resolutions based on different quality metrics. It has further shown that, when deep learning methods are used in an end to end compression system, there is flexibility in the distribution of complexity increases and these are not necessarily located at the encoder. Future work should focus on the complexity optimisation of the CNN to enable more efficient decoding process. 

\section*{Acknowledgment}

The authors acknowledge funding from the UK Engineering and Physical Sciences Research Council (EPSRC, project No. EP/M000885/1), and the NVIDIA GPU Seeding Grants. The authors would also like to thank Mr Di Ma for his contribution in developing the large video database for training the employed CNN.

\bibliographystyle{IEEEtran}
\footnotesize\bibliography{IEEEabrv,MyRef}

\begin{IEEEbiography}[{\includegraphics[width=1in,height=1.25in,clip,keepaspectratio]{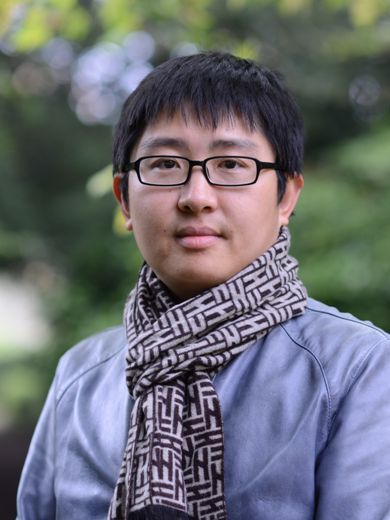}}]{Fan Zhang}
(M'12) received the B.Sc. and M.Sc. degrees from Shanghai Jiao Tong University (2005 and 2008 respectively), and his Ph.D from the University of Bristol (2012). He is currently working as a Senior Research Associate in the Visual Information Laboratory, Department of Electrical and Electronic Engineering, University of Bristol, on projects related to video compression and machine learning. His research interests include perceptual video compression, video quality assessment and machine learning based video coding.
\end{IEEEbiography}

\begin{IEEEbiography}[{\includegraphics[width=1in,height=1.25in,clip,keepaspectratio]{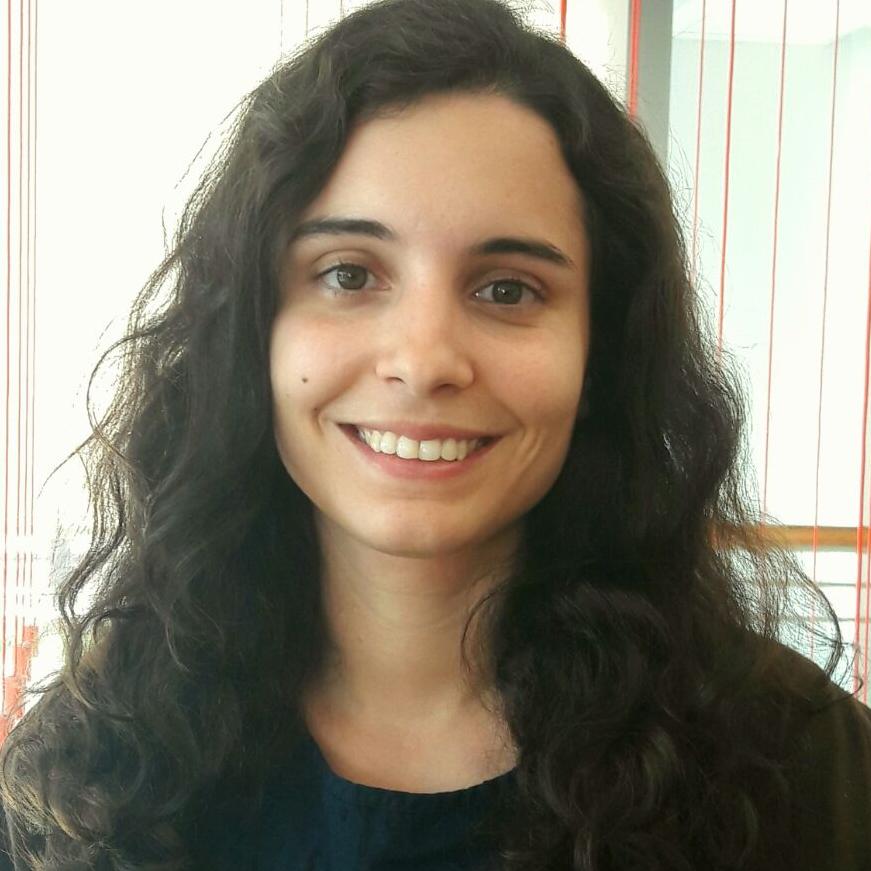}}]{Mariana Afonso}
received the B.S./M.S. degree in Electrical and Computers Engineering from the University of Porto, Portugal, in 2015 and a PhD in Electronic and Electrical Engineering from the University of Bristol, UK, in 2019. During her PhD, she was a part of the PROVISION ITN European Commission's FP7 Project, a network of leading academic and industrial organizations in Europe, working on the perceptual video coding. She also completed a secondment at Netflix, USA, in 2017. She is currently a Research Scientist in the Video Algorithms team at Netflix. Her research interests include video compression, video quality assessment, and machine learning.
\end{IEEEbiography}

\begin{IEEEbiography}[{\includegraphics[width=1in,height=1.25in,clip,keepaspectratio]{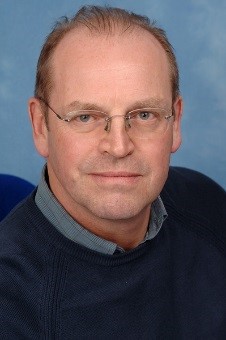}}]{David R. Bull}
(M'94-SM'07-F'12) received the B.Sc. degree from the University of Exeter, Exeter, U.K., in 1980; the M.Sc. degree from University of Manchester, Manchester, U.K., in 1983; and the Ph.D. degree from the University of Cardiff, Cardiff, U.K., in 1988.

Dr Bull has previously been a Systems Engineer with Rolls Royce, Bristol, U.K. and a Lecturer at the University of Wales, Cardiff, U.K. He joined the University of Bristol in 1993 and is currently its Chair of Signal Processing and Director of its Bristol Vision Institute. In 2001, he co-founded a university spin-off company,  ProVision Communication Technologies Ltd.,  specializing in wireless video technology.  He has authored over 450 papers on the topics of image and video communications and analysis for wireless, Internet and broadcast applications, together with numerous patents, several of which have been exploited commercially. He has received two IEE Premium awards for his work. He is the author of three books, and has delivered numerous invited/keynote lectures and tutorials. Dr. Bull is a fellow of the Institution of Engineering and Technology.
\end{IEEEbiography}

\end{document}